# Design of achromatic augmented reality visors based on composite metasurfaces

ELYAS BAYATI [1, *], ANDREW WOLFRAM [1], SHANE COLBURN [1], LUOCHENG HUANG[1], ARKA MAJUMDAR[1,2]

[1]Department of Electrical and Computer Engineering, University of Washington Seattle, WA-98195
[2]Department of Physics, University of Washington, Seattle, WA-98195
*Corresponding author: elyasb@uw.edu

A compact near-eye visor (NEV) system that can guide light from a display to the eye could transform augmented reality (AR) technology. Unfortunately, existing implementations of such an NEV either suffer from small field of view or chromatic aberrations. See-through quality and bulkiness further make the overall performance of the visors unsuitable for a seamless user experience. Metasurfaces are an emerging class of nanophotonic elements that can dramatically reduce the size of optical elements while enhancing functionality. In this paper, we present a design of composite metasurfaces for an ultra-compact NEV. We simulate the performance of a proof-of-principle visor corrected for chromatic aberrations while providing a large display field of view (>77° both horizontally and vertically), good see-through quality (>70% transmission and less than a wavelength root mean-square (RMS) wavefront error over the whole visible wavelength range), as needed for an immersive AR experience

## 1. Introduction

Augmented reality (AR) technology has recently attracted considerable attention and could revolutionize technologies ranging from applications in the military and navigation, to education and entertainment. Using these devices, we can effectively integrate computer-generated virtual information into the real world. AR can be implemented in various ways, including smartphone displays, head-up displays (HUDs), and head-mounted displays (HMDs) [1]. Among the various approaches, see-through head-mounted near-eye displays (AR glasses) have been of primary focus because of their potential to provide an extremely high sense of immersion. One of the most vital components in a head-mounted display (HMD) is a near-eye visor (NEV). Current implementations of NEVs can be broadly classified into two categories: **reflection-based**, where light from a display placed near the eye is reflected off a freeform visor to enter the eye [2-3]; and **waveguide-based**, where light is passed through a waveguide and projected to the eyes using a grating [4-6]. While reflective visors can provide higher efficiency and transparency, they are often bulky and suffer from a small eye-box, and in general need to be placed far away from the eyes to achieve a wide field of view (FOV). Waveguide-based NEVs can provide a much more compact form-factor with sufficient eye-box expansion, although they often have lower FOV and efficiency. They also suffer from strong chromatic aberrations in both the image passed through the waveguide as well as in see-through mode as the gratings are diffractive in nature. The core of a waveguide NEV consists of the input and output couplers. These can be either simple prisms, micro-prism arrays, embedded mirror arrays [7], surface relief gratings [8], thin or thick holographic optical elements (HOEs) [6], or diffractive optical elements (DOEs) [5]. The out-coupling efficiency of the gratings is, however, low, making the energy-efficiency of these waveguide-based visors poor relative to reflection-based visors. A closer look at reflective visors reveals that the trade-off between size and FOV of the NEV primarily originates from the fact that current NEVs rely on geometric reflections and refractions to bend light.

In recent years, sub-wavelength diffractive optics, commonly known as metasurfaces, have emerged as a versatile candidate to create ultra-thin, flat optical elements [9]. These metasurfaces are quasiperiodic arrays of subwavelength optical antennas that can modify the phase, amplitude, or polarization of an incident optical wavefront. This enables the creation of arbitrary optical surfaces, including those for freeform optical elements [10]. Unlike conventional diffractive optics, the subwavelength scatterers in a metasurface preclude higher-order diffraction, resulting in higher efficiency as all the light can be funneled into the zeroth order [11]. As diffraction can bend light by an angle more than reflection and refraction, it is possible to bring a metasurface closer to the eye, while maintaining a wide FOV. Recently, the first silicon metasurface freeform visor was designed by our group [12], providing a large field of view for virtual reality (VR) applications (77.3° both horizontally and vertically) when placed only 2.5 *cm* away from the eye. This metasurface visor, however, severely distorts the real-world view, has strong chromatic aberrations and is not transparent for AR applications. In addition to our group, several other metasurface-based visor designs have been reported in the literature for achieving large FOV, lighter designs and better see-through quality, including designs based on combinations of a metalens and dichroic mirrors

[13-14], Pancharatnam-Berry phase metalenses [15], and a compact light engine based on multilayers-metasurface optics [16].

In this paper, we propose a metasurface freeform optics-based NEV, that will circumvent real-world distortions and eliminate chromatic aberrations, as needed for an immersive AR experience. For this purpose, we propose to use a composite metasurface: one of the metasurfaces reflects light from the display to project the virtual world to the user's eye while the second metasurface circumvents the distortion of the real-world due to the first metasurface. The first metasurface is designed to have multi-chrome operation, which two metasurface together preserves the color information in transmission. The resultant display FOV of the composite metasurface NEV is more than 77° both horizontally and vertically which is better than current existing AR visors and also it has acceptable see-through quality over the visible range (less than a wavelength RMS wavefront error).

## 2. Paired Phase Masks

First, we design and model the single metasurface NEV that can guide light from a display at the HMD to the human eye via ray optical simulation (ZEMAX-OpticStudio). The size of the visor is assumed to be 4 *cm*×4 *cm* to maintain a compact form-factor comparable to that of ordinary sunglasses while still having a large field of view when placed only 2.5 cm away from the eyes. The display is placed between the visor and the eye: 1cm away and 1.5 cm upwards from the visor with an angle of 45° with respect to the optical axis. The display is initially assumed to be monochromatic (540 nm). The eye model which we use here for simulation is the widely accepted model proposed by Liou & Brennan [17]. Here, it is assumed that the eye is looking out through an optical system so the retina is the image surface. We model the visor phase-mask with Zernike standard polynomials [18]. We find that the first 6 Zernike terms converge in a reasonable time and are sufficient to guide light from the display to the human eye in an acceptable manner. A schematic of a ray tracing simulation from our reflective metasurface is shown in Fig. 1A. The first metasurface (near the eye) will be partially reflective, and the design principle will follow the ray tracing approach of sending light from each point on the display to the eye. The virtual display FOV from our simulation is estimated to be 77.3° along both vertical and horizontal directions. Fig. 1B illustrates the reflective phase-mask obtained from the Zernike standard polynomials from the first metasurface of our visor.

After designing the single metasurface near-eye visor, we design a pair of metasurfaces, where the second metasurface (located further from the eye) corrects any distortion of the real-world scene caused by the first metasurface. The gap size between two metasurfaces is assumed to be 500 μm which is fixed to facilitate future fabrication on both sides of a glass wafer. The corrective phase-mask of the second metasurface is also designed using Zernike standard polynomials. This phase mask is designed to minimize the RMS wavefront error [19] for light transmitted through two metasurfaces to ensure good see-through quality. Without using the corrective phase-mask the RMS wavefront error for light on the retina after passing through both metasurfaces is 14.8λ. Such wavefront errors represent the see through-quality, and with greater than wavelength rms error leads to a highly distorted view. However, after using the second metasurface to correct the distortion from the first metasurface, the RMS wavefront error is reduced to 0.9λ. We also calculated the RMS wavefront error of a freeform visor developed in [20] to compare to our see-through quality. The RMS wavefront error of the freeform visor is 1.17λ, which is better than a single metasurface but worse than our corrected composite metasurface visor.

Fig. 1 shows the optimized metasurface NEV phase profiles, where the first metasurface (Fig. 1B) distorts the light field, but in conjunction with the second metasurface (Fig. 1C), the optical wavefront error is minimized. Once the phase-functions are optimized, we implement them using metasurfaces, which is explained in more detail later in the paper.

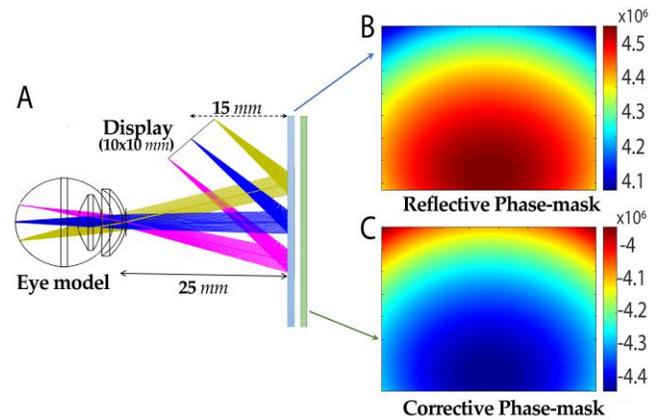

Figure 1. Metaform Visor (A) Schematics of metasurface NEV and its ray tracing simulation in ZEMAX. (B) Reflective phase-mask from first metasurface of our visor. (C) Corrective phase-mask for second metasurface of our visor. The size of the visor is assumed to be 4 cm×4 cm.

We also evaluate the modulation transfer function (MTF) and the grid distortion cause by the NEV to estimate the image quality. The MTF of the system along the tangential and sagittal plane is shown in Fig. 2A., exhibiting a value greater than 0.3 at 33 cycles/mm, which is a sufficient resolution for a human visual system [4] and satisfying the Nyquist resolution for a 15μm pixel size. The grid distortion is less than 5.9% which is sufficient for human intelligibility [4] (Fig. 2B.). Additionally, we simulate an image of the green crossbars (based on our operating wavelength design) in Zemax to evaluate the see-through quality before and after using the second phase-mask, as shown in Fig. 2C. The projected image of the crossbars is shown in the right side of Fig. 2C, which is the image reproduced on the retina after passing through the phase-masks and the eye model. The image distortion is clearly observed without our corrective phase-mask.

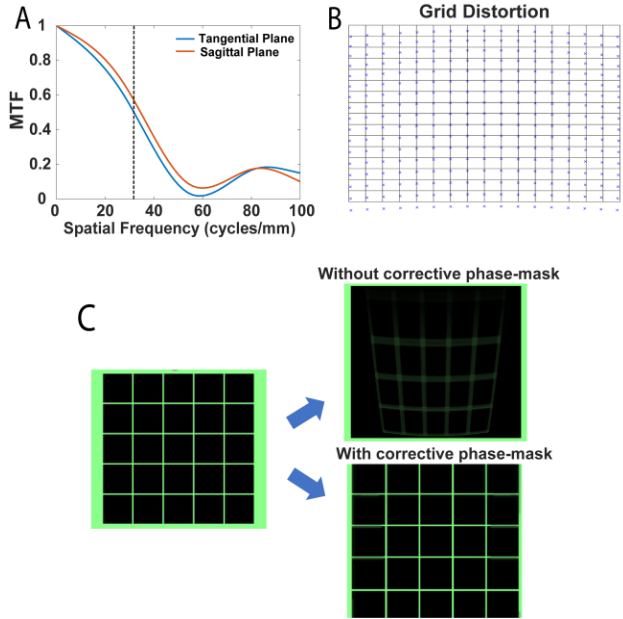

Figure 2. (A) MTF of the NEV calculated in ZEMAX on tangential and sagittal plane. (B) Grid distortion of the NEV corresponds to 10mm × 10mm display size. Highest calculated distortion is 5.9% at the corners. The corresponding operation wavelength here is 540 nm. (C) Image simulation of the green cross bars passing through NEV using Zemax. The left figure is the original image that is seen in real world. The right figure is the simulated image as seen by the person using NEV before and after adding corrective phase-mask.

## 3. Metasurface Implementation

So far, we assumed the NEV to be made of bilayer, ideal phase masks. In practice, these masks will be implemented using metasurfaces. The proposed stack of metasurfaces are shown in Fig. 3A. The partially reflective metasurface is designed by placing silicon nitride nano-scatterers on a partially reflective four-layer distributed Bragg reflector (DBR) [21]. The resulting phase and amplitude of the reflected light as a function of pillar diameter (duty cycle) are shown in Fig. 3B. Note that the amplitude of the reflected light is ~30%, whereas the phase covers the full 0-2π range. The second metasurface will be a transmissive one, whose transmission parameters are shown in Fig. 3C. The sharp resonances observed in the phase and amplitude correspond to guided mode resonances and are excluded when designing the final metasurface by selecting pillar diameters off resonance. The design process involves selecting the appropriate spatial phase profile for the specific optical component, arranging the scatterers on a subwavelength lattice, and spatially varying their dimensions. Inset figure of Fig. 3B. and 3C. shows cylindrical post formation including their substrates. The DBR structure in Fig. 3B. is designed for 540 nm wavelength so their thickness is 67.5 nm for SiN and 90 nm for oxide. The metasurface periodicity $p$ and thickness $t$ for both metasurfaces here are set to 443 nm and 700 nm respectively. All the simulations that we have done so far were for normal incident angle as shown in Fig. 3. However, based on our design schematic and display position in Fig. 1., our reflective metasurface will experience an angle of incidence from 40° to 50°. It is necessary for us to get the whole 0-2π phase coverage at these angles as well which is presented in more detail in Appendix A.

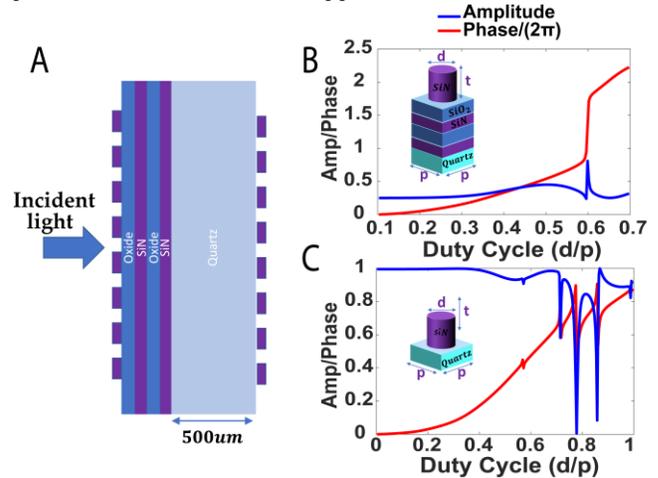

Figure 3. Metasurface (MS) implementation of the phase-masks: (a) schematic of the metasurface stack; the simulate phase and amplitude response of the (B) light reflected from the first metasurface and (C) light transmitted through the second metasurface. These plots are calculated using rigorous coupled wave analysis.

## 4. Multi-wavelength Metasurface

One of the major issues facing metasurfaces is their strong chromatic aberrations. For example, a metalens will produce a focused image at one wavelength and an unfocused image at another [22]. A successful AR visor, however, must produce focused images at multiple wavelengths, specifically red, green, and blue in order to display images for all perceivable colors. Much work has been done to produce multiwavelength achromatic metalenses (MAM) including multiplexing different functionalities in a single metasurface [23] or dispersion engineering [24], but until recently previous methods either have not been shown to work in the visible spectrum or have limitations on device size and the phase shift that can be imparted. Furthermore, such approaches have worked well for high-contrast scatterers (like silicon scatterers working in the infrared with refractive index n~3.5), as the light remains tightly confined inside the scatterer, and thus the coupling between different scatterers is minimized. Unfortunately, our application at visible wavelengths precludes silicon as the material of choice, and we have to rely on silicon nitride (n~2) and silicon dioxide (n~1.5), both of which are transparent and low index materials relative to silicon. The method described in Ref. [25], however, showed that it is possible to produce a metalens that operates at discrete wavelengths with large differences in the required phase shift at each wavelength. Our metasurface design employs this same method and demonstrates that guided mode resonances (GMR) can give rise to numerous 0-2π phase shift cycles observed for the duty cycle range for pillars using silicon nitride.

Our metasurface visor has a target phase profile at each wavelength that is position dependent. The pillar at any position on the visor must match the desired phases of all three wavelengths (R (700 nm), G (540 nm), B (460nm)) at that position simultaneously. Typically, for a monochromatic metasurface, when the phase shift of a pillar is chosen, the phase that it imparts at other wavelengths is then determined as well. Said another way, for a given phase shift, there is typically only one corresponding pillar size at each wavelength that imparts that phase shift. However, by scatterer design, we can have multiple pillar sizes at each wavelength that impart the desired phase shift. This enables us to better match phase profiles at different wavelengths as we can keep the phase shift at one wavelength fixed while selecting a different pillar size that may better match the phase profile at another wavelength. This one-to-many relationship between phase and pillar size is achieved by choosing pillar parameters such that the imparted phase shift spans 0-2π multiple times over the range of possible pillar sizes (Fig. 4.). The greater number of 0-2π cycles at each wavelength, the better we can design a metasurface that matches the phase profiles of all wavelengths. This approach is aided by the occurrence of GMR, which cause sudden, abrupt phase shifts that increase the number of pillar sizes available to fulfill a target phase shift.

Using rigorous coupled wave analysis (RCWA) [26], reflection coefficients and phases for pillars ranging in diameter from 10 to 90 percent of the periodicity were calculated for each wavelength. The lattice periodicity $p$ and thickness $t$ at all wavelengths here are set to 443 nm and 1500 nm, respectively, to achieve multiple 0-2π cycles at all three wavelengths. Pillars with reflection coefficients below 0.3 at any wavelength were removed from consideration as we want to strike a balance between transmission and reflection to maintain a partially reflective visor. For each discretized position in a phase mask, the remaining pillars were assigned an error weight calculated as the square of the difference between the phase mask at that position and the pillar's phase shift at that phase mask's wavelength. The metasurface in our proposed structure was then designed by selecting pillars with the smallest cumulative error ($\epsilon$) across the three phase masks at each position, which is given by:

$$\epsilon(x,y,r) = \sum_{i=1}^{n} \sqrt{\left(\phi(x,y,\lambda_i) - f(r,\lambda_i)\right)^2}$$

Where $(x,y)$ is the position on the phase mask, $\lambda_i$ is the design wavelength, $r$ is the pillar radius, and $n$ is the number of wavelengths (here $n = 3$). The desired phase is given by $\phi$ and the calculated phase shift for each pillar is given by $f$. The final radii distribution of our multi-wavelength reflective metasurface visor is shown in Fig. 4A. We emphasize that, in this method, we are primarily designing a metasurface that works at certain specific wavelengths, and true broadband operation is not expected. Hence, we need to rely on a display that supports only discrete wavelengths, such as a laser-based display [27].

We then validated our design of a reflective metasurface visor by importing the phase profiles that is provided by our final optimized scatterer distribution into the NEV structure via ray optic simulation which is shown in Fig.1. **Table 1** provides the modulation transfer function (MTF) for three colors at 33 cycles/mm (via Zemax). The MTFs for the multiwavelength metasurface are comparable to that of the monochrome design, exhibiting minimal degradation in performance. This clearly demonstrates that the multi-color metasurface does not degrade the performance of the visor compared to monochromatic metasurface one.

We also analyzed the bandwidth of the light-source for the display and we found that the MTF does not fall very quickly, and at least works for over a ~20 nm bandwidth. We define the bandwidth by the range of wavelengths over which the MTF exceeds 0.3 at 33 cycles/mm (Appendix B). Hence, in some cases, we also can use a narrowband light emitting diodes (LEDs), such as an organic LED, instead of a laser for the display.

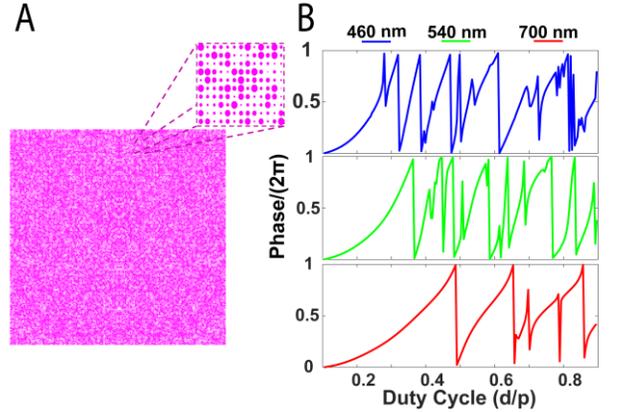

Figure 4. A) Final radii distribution of our multi-wavelength reflective metasurface visor using our optimization method. The diameter of nanopillars are changing from 46 nm to 398 nm. B) Phase of the reflected light through a scatterer using RCWA as a function of duty cycles of nanopillar at wavelengths of 460, 540, and 700 nm. As the pillar width changes, the corresponding phase undergoes multiple 0-2π phase cycles. The sharp phase jumps reflect the excitation of guided mode resonance (GMR).

**Table 1. AR visor performance in terms of MTF before and after RGB optimization**

| Wavelength ($nm$) | 460 | 540 | 700 |
|---|---|---|---|
| Single wavelength metasurface visor (cycles/mm) | 0.61 | 0.49 | 0.53 |
| multi-wavelength metasurface visor (cycles/mm) | 0.51 | 0.47 | 0.49 |

Once we finalize the multi-wavelength reflective visor design, we then optimize the design for the second metasurface at the central wavelength. We found that while the second metasurface is designed to negate the effect of the first metasurface, such negation happens over the full wavelength range simulated. As can be seen in Fig. 5, the RMS wavefront error is minimal over a broad optical bandwidth (simulated for different angles).

Additionally, we simulate an image of the white crossbars in Zemax to evaluate the see-through quality, as shown in Fig. 5.B. The projected image of the crossbars is shown in the right side of Fig. 5B, which is the image reproduced on the retina after passing through the metasurface layers and the eye model. We can clearly observe that the rainbow effect caused by chromatic aberration of these metasurfaces are negligible.

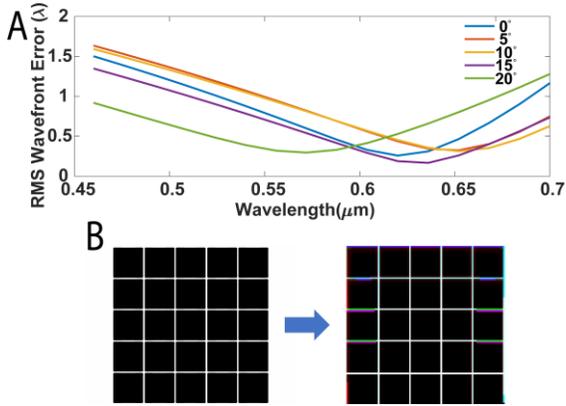

Figure 5. A) RMS wavefront error in visible range calculated for the transmitted light through the visor (light that passes through two metasurfaces) in five different incident angles. B) Image simulation of the white cross bars passing through NEV using Zemax. The left figure is the original image that is seen in real world. The right figure is the simulated image as seen by the person using NEV.

## 5. Discussion

The proposed composite metasurface visor can overcome the bulkiness, FOV limitations, chromatic aberrations, and see-through quality of existing NEVs. Using our proposed structure, we can make the eye-wear devices flat and ultra-thin and close to eye, while maintaining a large FOV for virtual world. However, there are some challenges with the current design, including small eye-box and large area fabrication. The eye-box for our current visor is 2.5mm×2.5mm (denoting exit pupil diameter without any expansion) which is lower than current waveguide-base visors [28], although there are some solutions to increase the eye-box by multiplexing different phase-masks into one metasurface [29].

Another remaining issue is fabrication. While the fabrication of the current design is challenging, fabrication methods already exist that show that such large-area fabrication using Deep Ultra-Violet (DUV) Lithography is possible with high throughput. We also have simulated the tolerance of the metasurfaces to lateral misalignment in terms of acceptable RMS wavefront error range (Appendix C).

We can also increase the efficiency of our current design by making our visor reflective only at certain display angles and totally transmissive at all other angles. Another limitation of the current proposed metasurface visor will be the vergence-accommodation conflict (VAC) which is common in current AR glasses. However, a recent study shows that by using a tunable focal length lens, the VAC can be reduced [30-31]. By integrating metasurface structures intro micro-electromechanical systems such tunability can be achieved which can mitigate the VAC [32-33].

## 6. Conclusion

Our work explored the possibility of using a metasurface to design and create a compact near-eye visor which provides a large field of view and also reasonable see-through quality for an immersive AR experience. The proposed metasurface visor does not suffer from chromatic aberrations, while providing a large display field of view (>77° both horizontally and vertically), and good (>70% transmission and no distortion) see-through quality.

The current visor includes two layers of metasurfaces which have different phase masks. The scatterers in the metasurface near the eye (which is reflecting the virtual world to the eye) are designed to have 30% reflectivity and multi-chrome behavior, whereas the scatterers in the metasurface further from the eye (which helps improve the see-through quality) are designed to have 100% transmission. The main idea behind our multi-chrome behavior is that we start with the desired phase profiles for each color, and then find the scatterer distribution that provide the phase profiles closest to the desired one.

In this work, we focused on design and simulation of the whole AR metasurface visor due to the complexity of design. Additionally, there are many other relevant factors for high-quality immersive AR experience including large eye-box, high efficiency, broadband operation and tunability (to mitigate VAC) which are not specified in our current architecture. However, we believe that the demonstrated multi-layer metasurface visor is capable of addressing those issues.

## 7. Appendices

### A. Angle dependency

All the RCWA simulations for the phase and amplitude calculation (in reflection and transmission mode) were performed under normal incidence as shown in Fig. 3. However, based on our design schematic and display position in Fig. 1., our reflective metasurface will experience an angle of incidence from 40° to 50°. It is necessary for us to get the whole 0-2π phase coverage at these angles as well. As it is shown in Fig. 6., at these angles we still get multiple wraps of 0-2π phase change which is ideal for our multi-wavelength design approach. Here, the operation wavelength for angle dependency simulation is 540 nm.

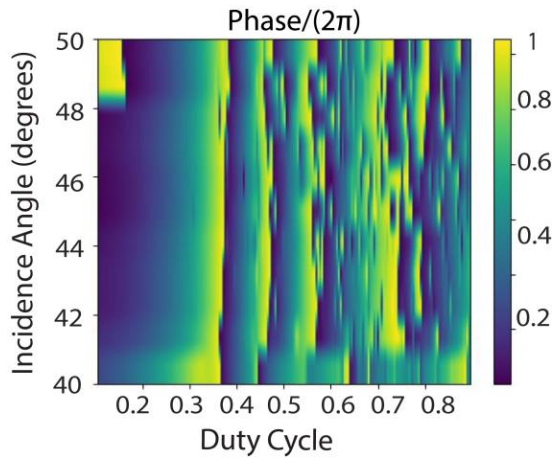

Figure 6. Angle dependency RCWA simulation from 0 to 2π for one wavelength (540 nm). Duty cycle is defined as the ratio of the post diameter to the periodicity. Parameters of pillars are the same for Fig.3.

### B. Bandwidth Tolerance

As we mentioned in section 4, we have primarily designed a metasurface that works at certain specific wavelengths, and true broadband operation is not expected here. However, we also analyzed the bandwidth of the light-source for the display and we found that the MTF does not fall very quickly, and at least works for over a ~20 nm bandwidth. We define the bandwidth by the range of wavelengths over which the MTF exceeds 0.3 at 33 cycles/mm (Fig. 7.). Hence, we can use an LED instead of a laser for the display.

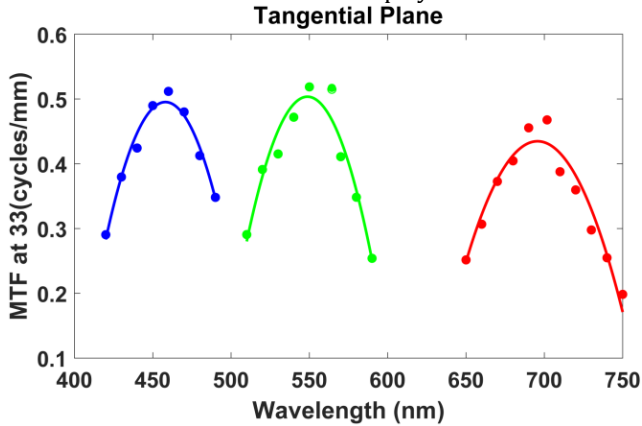

Figure 7. The MTF of the blue, green and red color for the reflective visor at 33 cycles/mm in tangential plane. MTFs above 0.3 at 33 cycle/mm are our evaluation point for acceptable bandwidth.

### C. Fabrication Tolerance

While the fabrication of the current design is challenging, fabrication methods already exist that show that such large-area fabrication using DUV Lithography is possible with high throughput as we discussed in section 5. One of the challenging tasks will be to align two metasurfaces during fabrication on both side of a quartz substrate. Here, we have simulated the tolerance of the metasurfaces to lateral misalignment. In this test simulation, the operation wavelength is 540 nm and the angle of incidence is assumed to be normal. We are looking for changes in RMS wavefront error in see through mode. We found out that up to 100μm misalignment, error remains below one wavelength, which is still close to our current RMS at normal incident angle.

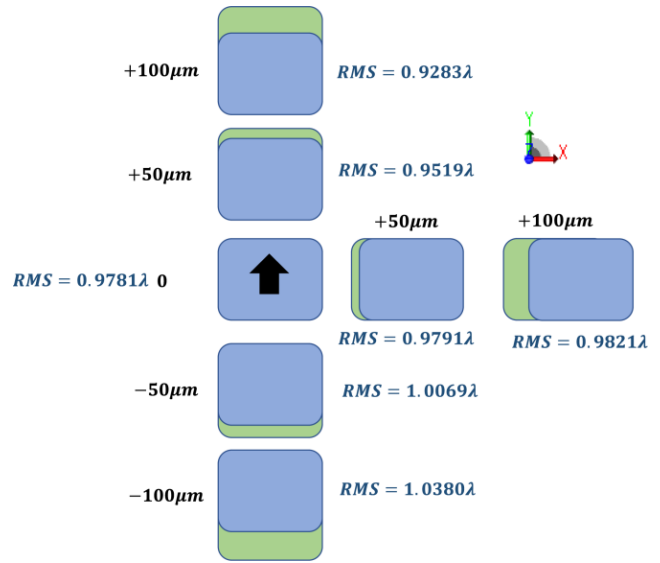

Figure 8. Analysis of lateral tolerance in terms of the RMS wavefront error in the transmitted light. Up to ~100μm misalignment, error remains below one wavelength. The operation wavelength here is 540nm and the angle of incidence is assumed to be normal. The blue box represents the first metasurface (Reflective phase-mask), and the green represents the second metasurface (Corrective phase-mask). The arrow mark shows us which side of the metasurface is up compared to our eye view.

**Funding.** The research is supported by the Samsung Global Research Outreach (GRO) grant, and the UW Reality Lab, Facebook, Google, and Futurewei.

**Acknowledgment.** We thank Optics studio (Zemax) corporation for the available ray tracing simulation and optimization method.

**Disclosures.** The authors declare no conflicts of interest.

### References

1. Azuma, R., Baillot, Y., Behringer, R., Feiner, S., Julier, S., & MacIntyre, B. (2001). Recent advances in augmented reality. IEEE computer graphics and applications, 21(6), 34-47.
2. Hu, X., & Hua, H. (2014). High-resolution optical see-through multi-focal-plane head-mounted display using freeform optics. Optics express, 22(11), 13896-13903.
3. Pan, J. W., Che-Wen, C., Huang, K. D., & Wu, C. Y. (2014). Demonstration of a broad band spectral head-mounted display with freeform mirrors. Optics express, 22(11), 12785-12798.
4. Yang, J., Twardowski, P., Gérard, P., & Fontaine, J. (2016). Design of a large field-of-view see-through near to eye display with two geometrical waveguides. Optics letters, 41(23), 5426-5429.

5. Liu, Z., Pan, C., Pang, Y., & Huang, Z. (2019). A full-color near-eye augmented reality display using a tilted waveguide and diffraction gratings. Optics Communications, 431, 45-50.
6. Yoshida, T., Tokuyama, K., Takai, Y., Tsukuda, D., Kaneko, T., Suzuki, N., ... & Machida, A. (2018). A plastic holographic waveguide combiner for light-weight and highly-transparent augmented reality glasses. Journal of the Society for Information Display, 26(5), 280-286.
7. Kress, B. C. (2019, July). Optical waveguide combiners for AR headsets: features and limitations. In Digital Optical Technologies 2019 (Vol. 11062, p. 110620J). International Society for Optics and Photonics.
8. Waldern, J. D., Grant, A. J., & Popovich, M. M. (2018, May). DigiLens switchable Bragg grating waveguide optics for augmented reality applications. In Digital Optics for Immersive Displays (Vol. 10676, p. 106760G). International Society for Optics and Photonics.
9. Yu, N., & Capasso, F. (2014). Flat optics with designer metasurfaces. Nature materials, 13(2), 139-150.
10. Zhan, A., Colburn, S., Dodson, C. M., & Majumdar, A. (2017). Metasurface freeform nanophotonics. Scientific reports, 7(1), 1-9.
11. Stork, W., Streibl, N., Haidner, H., & Kipfer, P. (1991). Artificial distributed-index media fabricated by zero-order gratings. Optics letters, 16(24), 1921-1923.
12. Hong, C., Colburn, S., & Majumdar, A. (2017). Flat metaform near-eye visor. Applied Optics, 56(31), 8822-8827.
13. Lee, G. Y., Hong, J. Y., Hwang, S., Moon, S., Kang, H., Jeon, S., ... & Lee, B. (2018). Metasurface eyepiece for augmented reality. Nature communications, 9(1), 1-10.
14. Lan, S., Zhang, X., Taghinejad, M., Rodrigues, S., Lee, K. T., Liu, Z., & Cai, W. (2019). Metasurfaces for near-eye augmented reality. ACS Photonics, 6(4), 864-870.
15. Moon, S., Lee, C. K., Nam, S. W., Jang, C., Lee, G. Y., Seo, W., ... & Lee, B. (2019). Augmented reality near-eye display using Pancharatnam-Berry phase lenses. Scientific reports, 9(1), 1-10.
16. Kamali, S. M., Arbabi, E., & Faraon, A. (2019, February). Metasurface-based compact light engine for AR headsets. In Optical Design Challenge 2019 (Vol. 11040, p. 1104002). International Society for Optics and Photonics.
17. Atchison, D. A., & Smith, G. (2000). Optics of the human eye. Butterworth-Heinemann.
18. Von F, Z. (1934). Beugungstheorie des schneidenver-fahrens und seiner verbesserten form, der phasenkontrastmethode. physica, 1(7-12), 689-704.
19. Shannon, R. R. (1995). Optical Specification. Handbook of Optics, 1, 35-1.
20. Cheng, D., Hua, H., & Wang, Y. (2012). U.S. Patent Application No. 13/318,864.
21. Sheppard, C. J. R. (1995). Approximate calculation of the reflection coefficient from a stratified medium. Pure and Applied Optics: Journal of the European Optical Society Part A, 4(5), 665.
22. Colburn, S., Zhan, A., & Majumdar, A. (2018). Metasurface optics for full-color computational imaging. Science advances, 4(2), eaar2114.
23. Arbabi, E., Arbabi, A., Kamali, S. M., Horie, Y., & Faraon, A. (2016). Multiwavelength metasurfaces through spatial multiplexing. Scientific reports, 6, 32803.
24. Chen, W. T., Zhu, A. Y., & Capasso, F. (2020). Flat optics with dispersion-engineered metasurfaces. Nature Reviews Materials, 1-17.
25. Shi, Z., Khorasaninejad, M., Huang, Y. W., Roques-Carmes, C., Zhu, A. Y., Chen, W. T., ... & Devlin, R. C. (2018). Single-layer metasurface with controllable multiwavelength functions. Nano letters, 18(4), 2420-2427.
26. Liu, V., & Fan, S. (2012). S4: A free electromagnetic solver for layered periodic structures. Computer Physics Communications, 183(10), 2233-2244.
27. Chellappan, K. V., Erden, E., & Urey, H. (2010). Laser-based displays: a review. Applied optics, 49(25), F79-F98.
28. Miller, J. M., De Beaucoudrey, N., Chavel, P., Turunen, J., & Cambril, E. (1997). Design and fabrication of binary slanted surface-relief gratings for a planar optical interconnection. Applied optics, 36(23), 5717-5727.
29. Kim, S. B., & Park, J. H. (2018). Optical see-through Maxwellian near-to-eye display with an enlarged eyebox. Optics letters, 43(4), 767-770.
30. Padmanaban, N., Konrad, R., Stramer, T., Cooper, E. A., & Wetzstein, G. (2017). Optimizing virtual reality for all users through gaze-contingent and adaptive focus displays. Proceedings of the National Academy of Sciences, 114(9), 2183-2188.
31. Wang, Y. J., Lin, Y. H., Cakmakci, O., & Reshetnyak, V. (2020). Varifocal augmented reality adopting electrically tunable uniaxial plane-parallel plates. Optics Express, 28(15), 23023-23036.
32. Han, Z., Colburn, S., Majumdar, A., & Bohringer, K. (2020). MEMS-actuated Metasurface Alvarez Lens. arXiv preprint arXiv:2001.07800.
33. Arbabi, E., Arbabi, A., Kamali, S. M., Horie, Y., Faraji-Dana, M., & Faraon, A. (2018). MEMS-tunable dielectric metasurface lens. *Nature communications*, *9*(1), 1-9.